\documentclass{iaus}
\usepackage{graphicx}

\title[The Dynamical Evolution of Be star disks] 
{The Dynamical Evolution of Be star disks}

\author[X. Haubois \& al.]   
{Xavier Haubois$^1$, 
 Alex C. Carciofi$^1$,
Atsuo T. Okazaki$^2$, \\
 \and Jon E. Bjorkman$^3$
 }

\affiliation{$^1$Instituto de Astronomia, Geof\'{i}sica e Ci\^{e}ncias Atmosf\'{e}ricas, Universidade de S\~{a}o Paulo, Rua do Mat\~{a}o 1226, Cidade Universit\'{a}ria, S\~{a}o Paulo, SP 05508-900, Brazil\\ email: {\tt xhaubois@astro.iag.usp.br} \\[\affilskip]
$^2$Faculty of Engineering, Hokkai-Gakuen University, Toyohira-ku, Sapporo 062-8605, Japan \\[\affilskip]
$^3$University of Toledo, Department of Physics \& Astronomy, MS111 2801 W. Bancroft Street Toledo, OH 43606, USA
}

\pubyear{2010}
\volume{xxx}  
\pagerange{119--126}
\jname{IAUS 272 Active OB stars - structure, evolution, mass loss, and critical limits}
\editors{A.C. Editor, B.D. Editor \& C.E. Editor, eds.}
\begin{document}

\maketitle

\begin{abstract}
We present a novel theoretical tool to analyze the dynamical behaviour of a Be disk fed by non-constant decretion rates. It is mainly based on the computer code HDUST, a fully three-dimensional radiative transfer code that has been successfully applied to study several Be systems so far, and the SINGLEBE code that solves the 1D viscous diffusion problem. We have computed models of the temporal evolution of different types of Be star disks for different dynamical scenarios. By showing the behaviour of a large number of observables (interferometry, polarization, photometry and spectral line profiles), we show how it is possible to infer from observations some key dynamical parameters of the disk.

\end{abstract}

\firstsection 

\section{Presentation of the simulations}
In order to analyze the observational signatures of different dynamical
scenarios, we use the SINGLEBE code (\cite{2007ASPC..361..230O}) that computes the temporal evolution of the 
surface density for a given stellar mass loss history and an $\alpha$ viscosity parameter (\cite{1973A&A....24..337S}) of the disk.
We then use the surface density at a given time as an input for a three-dimensional non-LTE Monte Carlo code called HDUST (\cite{car06}). By repeating this for different epochs of the disk evolution, we can follow the evolution of several observables (SED, polarization, images, etc).


A wide range of observables can be derived from the HDUST
simulations. By means of a comparative work, they can
unveil the observational features of a specific dynamical scenario
occurring in a viscous disk. The final objective of these quantities is to
be analyzed and compared to real polarimetric, photometric,
spectroscopic or interferometric observations. This method has the
potential to allow one observer to unambiguously infer the size, the $\alpha$ parameter and the mass loss history of the observed system.

\section{Some results}

\begin{figure}[h!]
\begin{center}
 \includegraphics[width=4.9in]{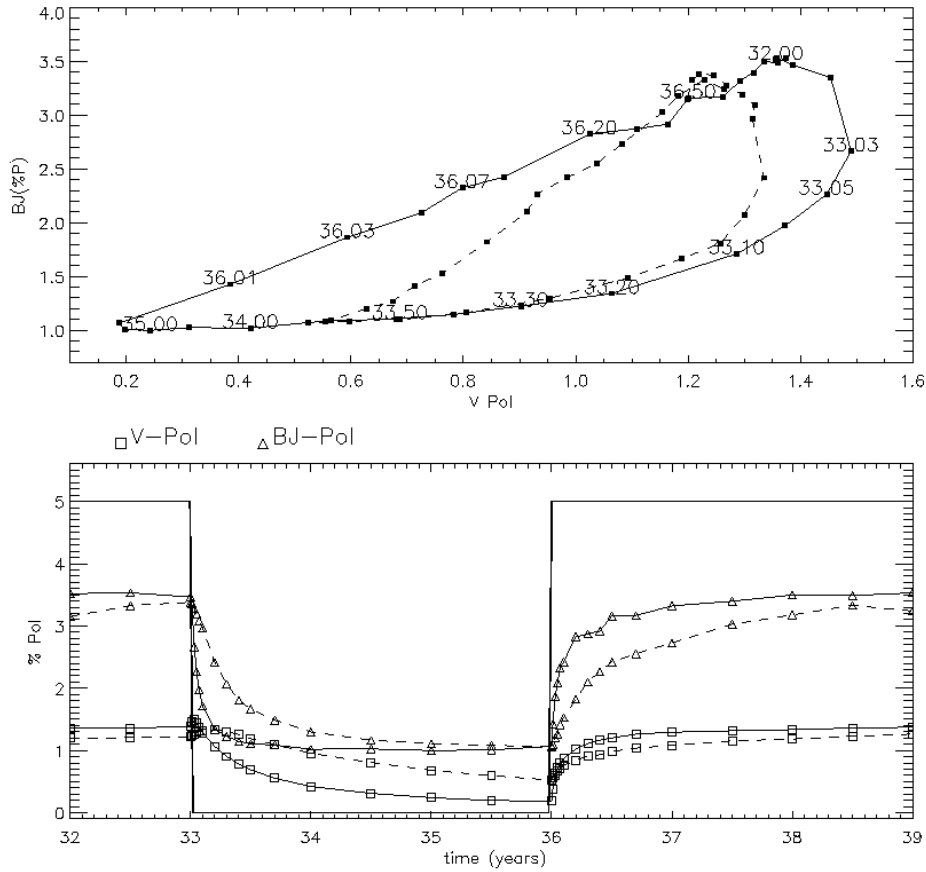} 
 \caption{Upper panel: polarization change across the Balmer Jump (BJ) vs V band polarization for two $\alpha$ parameter values (0.1 in dashed line, 0.7 in full line) and at an inclination angle of 70 degrees. Epochs are marked and counted in years, the
correlation loops move clockwise. Lower panel: temporal evolution of these two quantities. The thickest line shows the mass loss history arbitrarily scaled to the range of the graphic. It has been turned on and off every 3 years from 0 to 39 years. That figure depicts the 32-39 years period.}

   \label{fig1}
\end{center}
\end{figure}

As preliminary results, we present some observables and their correlation which clearly show different temporal behaviour depending on the $\alpha$ parameter (Fig.÷\ref{fig1}).
We present results for a dynamical scenario in which 3 year long outbursts are separated by 3 year long quiescent phases.
The influence of $\alpha$ is understood by considering that the larger $\alpha$, the faster the viscous diffusion occurs.

\section{Conclusion and Perspectives}

These simulations can predict the observational signatures of dynamical scenarios by generating a wide range of observables.
Some of these observables are thus powerful tools to estimate some key parameters of observed Be systems such as the $\alpha$ viscosity parameter. This work is currently ongoing and more complex dynamical scenarios have to be investigated (\cite{PapI}).  Moreover, the results of these studies being more easily understandable with videos, we plan to make them accessible to the community in a near future through a dedicated website (http://www.astro.iag.usp.br/beacon/).

\end{document}